\documentclass[a4paper]{panl}
\pdfoutput=1
\usepackage{cite}
\usepackage{wrapfig}
\usepackage{graphicx}
\usepackage{amssymb}
\usepackage{amsfonts}
\usepackage{amsmath}
\usepackage{longtable}
\usepackage{rotating}
\usepackage{lscape}
\usepackage{epsfig}
\usepackage{multirow}

\usepackage{lineno}

\originalTeX

\newcommand{\tit}[1]{}

\renewcommand{\d}{{\rm d}}

\newcommand{\blambda}{\bar\lambda}
\newcommand{\brho}{\bar\rho}
\newcommand{\tr}{{\rm tr}\,}
\renewcommand{\C}{{\bar \lambda}}
\renewcommand{\!}{\negthinspace}

\def\e{{\,\rm e}}
\def\be{\begin{equation}}
\def\ee{\end{equation}}
\def\beq{\begin{equation}}
\def\eeq{\end{equation}}
\def\bea{\begin{eqnarray}}
\def\eea{\end{eqnarray}}

\newcommand{\eps}{\varepsilon}
\newcommand{\non}{\nonumber \\*}

\newcommand{\vp}{\varphi}
\newcommand{\p}{\partial}
\newcommand{\bp}{\bar\partial}
\newcommand{\tbl}[1]{#1}
\newcommand{\tre}[1]{#1}
\newcommand{\tbr}[1]{#1}

\newcommand{\tma}[1]{#1}
\newcommand{\tcy}[1]{#1}
\newcommand{\tgr}[1]{#1}

\begin{document}

\title{What Quantum Strings can tell us  about\\ Quantum Gravity\footnote{Talk at the
"International Conference on Quantum Field Theory, High-Energy Physics, and Cosmology" Dubna July 18--21, 2022}
\vspace*{3mm}}
\maketitle
\authors{Yuri Makeenko
}
\setcounter{footnote}{0}
\from{Institute of Theoretical and Experimental Physics, Moscow}
\from{E-mail: makeenko@itep.ru}


\begin{abstract}
I describe the recent progress in resolving two problems  of nonperturbative bosonic string inherited from 1980's.
Both the lattice and KPZ-DDK no-go theorems can be bypassed thanks to 
specific features of the theory with diffeomorphism invariance.
\end{abstract}

\vspace*{6pt}


\label{sec:intro}
\section{{\bf Introduction}}

This Talk is based on the recent papers \cite{Mak,AM} written partially in collaboration with
Jan Ambj\o rn. Accurate references to the cited results  of 
other authors can be found there.
While the papers are relatively new, the problems they are devoted are {inherited from 1980's}
and were formulated as two no-go theorems for string existence.

\tit{Two no-go theorems for string existence}

1) {The non-perturbative}  regularization of the Nambu-Goto string by a hypercubic lattice 
 {does not} scale to a continuum string for the embedding space dimension 
 $d\geq2$ as found by {\it Durhuus, Fr\"ohlich, Jonsson (1984)}.
Analogously the  regularization of the  Polyakov string by dynamical triangulation scales to
continuum only for $d\leq 1$ but does not for $d>1$  as shown by 
{\it Ambj\o rn, Durhuus (1987)}.

2) The {string susceptibility} index  of  (closed) {Polyakov's string}, calculated by
{\it Knizhnik-Polyakov-Zamolodchikov (1988), David (1988), Distler-Kawai (1989)} 
using the conformal theory technique,
 is {not real}  for $1<d<25$
$$
\gamma_{\rm str}=(1-{h})\frac{d-25-\sqrt{(d-1)(d-25)}}{12} +2
\qquad \fbox{{genus} $h$}
$$

The possible solutions of these two problems are described in my talk and {rely on
subtleties} in quantum field theory  enjoying {diffeomorphism invariance} like
{Strings(!)} and {Gravity(?)}:  

{1) The continuum limit is not as in usual quantum field theory: the Lilliputian scaling regime 
versus an infinite correlation length}.

{2) The Nambu-Goto and Polyakov strings are told apart} by
{higher derivative terms in the emergent action which are classically suppressed
with the UV cuttoff $\Lambda\to\infty$
 as $\sim \Lambda^{-2}$  but revive  quantumly as $\Lambda^{-2}\times \Lambda^2 \sim 1$. }

\label{sec:preparation}
\section{{\bf Mean-field ground state of bosonic string}}

\tit{Nambu-Goto versus Polyakov strings}

The action of the {Polyakov string} is
{quadratic} in the target-space coordinate
$X^\mu$ and has an {independent} metric tensor $\rho_{ab}$.
{The Nambu-Goto string} can also be written like it by introducing 
 {the Lagrange multiplier} $\lambda^{ab}$.
 The two string formulations are expected to be equivalent as shown classically by
{\it Polyakov (1981)} and at one loop by {\it Fradkin, Tseytlin (1982)}.
The general argument is given in the book by {\it Polyakov (1987)}.

We consider a {closed bosonic string} winding once around {compactified}  dimension
of circumference $\beta$ and propagating through an (Euclidean) time $L$ 
{(topology of a cylinder or a torus)}. 
There are no tachyonic states if $\beta$ is {large enough}.
Choosing  the world-sheet parameters  $\omega_1,\omega_2 \in\omega_L\times \omega_\beta$ 
inside a rectangle,
the {classical ground state}  $X^\mu_{\rm cl}$ and $\rho^{ab}_{\rm cl}$  is usual and
$\lambda^{ab}_{\rm cl}
=  \rho^{ab}_{\rm cl} \sqrt{\rho_{\rm cl}}$ simplifies to
$\lambda^{ab}_{\rm cl}=\delta^{ab}$ in the {conformal gauge} $\rho_{ab}=\rho \delta_{ab}$ for 
$\omega_L/\omega_\beta=L/\beta$.

\tit{Induced (or emergent) action}

Let us do the {Gaussian} path
integral over $X^\mu_{\rm q}$ by splitting
$X^\mu=X^\mu_{\rm cl} +X^\mu_{\rm q}$:
$$
S_{\rm ind}= K_0 \int \d^2\omega\,\sqrt{ \rho} 
+\frac{K_0}2 \int \d^2\omega\, \lambda^{ab} \left( \partial_a X_{\rm cl } \cdot \partial_bX_{\rm cl } -\rho_{ab} \right) + \frac{d}{2}  \tr \log {\cal O},$$
where the
{operator ${\cal O}= -\frac1 {\sqrt{\rho} }  \partial _a \lambda^{ab} \partial_b$ 
reproduces  the Laplacian $\Delta$ for
$\lambda^{ab}=\rho^{ab} \sqrt{\det \rho}$}.
An {additional}  {ghost determinant} also emerges as usual
in the conformal gauge.
The action is called 
{induced} (or {emergent}). It coincides with the {effective} 
action for {smooth} fields.

Two-dimensional  determinants diverge and have to be {regularized}.
\tit{Regularization of the determinants}
For Schwinger's {proper-time} regularization
the integrals over $\tau$ are simply cut from below at $a^2=1/4\pi \Lambda^2$.
We use instead 
{Pauli-Villars'} regularization  by {\it Ambj\o rn, Y.M.~(2017)}
$$
\det({\cal O})\big|_{\rm reg}\equiv\frac{\det({\cal O})\det({\cal O}+2M^2)}{\det({\cal O}+M^2)^2}
$$
when
$$
 \tr\log{\cal O}\big|_{\rm reg}= -
\int_{0}^\infty \frac{\d \tau}{\tau} \,\tr \e^{-\tau{\cal O}}
\left(1-\e^{-\tau M^2}\right)^2, \quad ~~\Lambda^2 = \frac{M^2}{2\pi}\log 2
$$
is {convergent} at finite {regulator mass} $M$ and divergent as $M\to\infty$.

For  {Pauli-Villars'} regularization a beautiful diagrammatic  technique applies and 
the det's can be {exactly}
computed for certain metrics by the
 {Gel'fand-Yaglom} technique  to compare with the {Seeley} expansion
which starts with the term  $1/\tau$
in  two dimensions.
For $\tau\sim 1/\Lambda^2$ the higher terms are suppressed as $R/\Lambda^2$
for {smooth} fields {but revive if they are not}.

\tit{Mean-field ground state}

It is easy to compute the effective action  $S_{\rm eff}$ for {diagonal} and {constant}
$\lambda^{ab}=\blambda \delta^{ab}$ and $\rho_{ab} =\brho\delta_{ab} $.
Omitting the boundary terms  for $L\gg\beta$,
the minimum of $S_{\rm eff}$ is reached at the
{quantum ground state}
$$
\blambda=\frac{1}2 \left(1+\frac{\Lambda^2}{K_0} +
\sqrt{\left(1 +\frac{\Lambda^2}{K_0}\right)^2 -\frac{2d\Lambda^2}{K_0}}\right),\qquad
\brho 
\propto \frac \C {\sqrt{ \Big(1 +\frac{\Lambda^2}{K_0}\Big)^2 -\frac{2d\Lambda^2}{K_0}}}
$$
as found by  {\it Ambj\o rn, Y.M. (2017)}.
The minimization over  $\omega_\beta/\omega_L$ is
also needed at the saddle point.

The approximation describes a {mean field}  which takes into account 
an infinite set of pertubative diagrams 
about the classical vacuum. 
Then $\lambda^{ab}$ and $\rho_{ab}$ do not fluctuate 
in the mean-field approximation 
which becomes {exact}  at large $d$.

It is like the two-dimensional $O(N)$ sigma model at large $N$, where the Lagrange multiplier
does not fluctuate (summing the bubble graphs).
The large-$N$ vacuum is very close to the physical vacuum even for $N=3$.

The square root in $\blambda$ and $\brho$ is well-defined for ${d \geq 2 }$ if
$$
K_0>K_*=\left(d-1+\sqrt{d^2-2d} \right)\Lambda^2 \stackrel{d\to\infty}\to
2d \Lambda^2
$$
The
perturbation theory is recovered by expanding in $1/K_0\sim \hbar$.
Then $\C$ ranges  between 1 ({the classical value}) and  the ({quantum}) value
$$
\C_*=\frac 12\left(d -\sqrt{d^2-2d} \right) \stackrel{d\to\infty}\to \frac 12
$$

A natural question is as to why the minimum of the action  is reached at constant $\brho$
and $\blambda$.  I shall confirm it in Sect.~\ref{s:6} by showing a stability of
this ground state under fluctuations.

\section{{\bf Two scaling regimes: Gulliver's vs.\ Lilliputian}}

\tit{Lattice-like scaling limit (Gulliver's)}

The mean-field {ground state energy} is like found by {\it Alvarez (1981), Arvis (1983)} 
$$
E_N(\beta)=K_0 \C \sqrt{\beta^2+\frac{1 }{K_0 \C} \left(-\frac{\pi (d-2) }3 +8N\right)}
$$
and does {not} scale because $K_0>K_*\sim \Lambda^2$ for $\C$ to be real ($>\C_*$).
Let us choose 
$\beta^2$ slightly larger $\beta^2_{\rm min}=\frac{\pi(d-2)}{3K_*\C_* }$ 
for not to have a tachyon. It is clear that  the ground-state energy $E_0(\beta)$ 
can be made finite
by fine tuning $\beta$.

This scaling does not exist for excited states 
and thus is {particle-like} similar to the lattice regularizations of a string, where 
{only the lowest mass scales to finite while excitations scale to infinity},
reproducing the results by
{\it Durhuus, Fr\"ohlich, Jonsson (1984),   Ambj\o rn, Durhuus (1987)}.

\tit{Lilliputian string-like scaling limit}

Let us ``{renormalize}'' the {units of length}
$$
L_R=\sqrt{ \frac{\C}{\C-\C_*}}\, L,\qquad 
\beta_R=\sqrt{ \frac{\C}{\C-\C_*}} \,\beta
$$
to obtain a {finite} effective action
$$
S_{\rm mf} = K_R \; L_R \sqrt{ \beta_R^2 - \frac{\pi (d-2)}{3 K_R}},\qquad K_R = 
K_0 (\C-\C_*)
$$
The {renormalized string tension} $K_R$ scales to {finite} if 
$ K_0\to K_*$
reproducing the {Alvarez-Arvis}  spectrum of the continuum string.
The {average area} is also finite
recovering  the minimal area for large $\beta_R$.  

\tit{The Lilliputian world}

The Lilliputian scaling regime  is 
analogous to the zeta-function regularization except for the {nonlinearities}, but
the bare length in target space 
 is of order of the {cutoff}.  This is why it was called {Lilliputian}.
Such a scaling exists because  $\brho\to\infty$ 
as $K_0\to K_*$.

For this reason the cutoff in parameter space is
{\fbox{$ \Delta \omega=1/(\Lambda \sqrt[4]{g})$}} which 
fixes the maximal number of  modes in the {mode expansion} to be
$
n_{\rm max}\sim\Lambda  \sqrt[4]{g} \omega_\beta
$.
{Classically} $\sqrt[4]{g} \omega_\beta=\beta$  but  {quantumly}  $\sqrt[4]{g} \omega_\beta\propto
\frac{\sqrt{K_0}\,\beta }{\sqrt{K_R} }$ 
{is {much larger}.}

The Lilliputian scaling describes
continuum because infinitely smaller distances can be probed {$\Longrightarrow$}
  {classical music can be played on the Lilliputian strings.}
Gulliver's tools are too coarse to resolve the {Lilliputian world}. 
This is why the lattice string regularizations of 1980's never reproduce
 {canonical quantization.}

\section{{\bf Instability of the classical ground state}} 
\tit{Semiclassical energy}

The semiclassical (or one-loop) correction 
due to \tbr{zero-point fluctuations} was first computed by {\it Brink, Nielsen (1973)}
$$
S_{1l}=\left[K_0-\frac{(d-2)}2 \Lambda^2 \right] L\beta -\frac{\pi (d-2) L}{6\beta} 
$$
To make it finite, it is custom to introduce the \tbl{renormalized string tension}
$$
K_R=K_0-\frac{(d-2)}2 \Lambda^2 
$$
which is kept \tre{finite} as $\Lambda\to\infty$. Then it is assumed  it works order by
order of the perturbative expansion about the classical ground state, so  $K_R$ can be
made finite by fine tuning $K_0$.

We see however that the mean-field action
$S_{\rm mf}$
\tre{never vanishes} with changing $K_0$ \tcy(except for $\beta=\beta_{\rm min}$). 
Thus the \tcy{one-loop
correction simply lowers} for $d>2$ the energy of the classical ground state which 
may indicate its \tma{instability}.

\tit{Effective potential}

To check a (in)stability of the ground state, let us
add the source term 
$$
S_{\rm src}=\frac {K_0}2 \int \d^2 \omega \, j^{ab} \rho_{ab}
$$
and define the field
$
\rho_{ab}(j)
$
in the usual way.
Minimizing 
for constant $j^{ab}=j \delta ^{ab}$, {\it Ambj\o rn, Y.M.\ (2017)} found
the ``effective potential''  
in the mean-field approximation 
$$
\Gamma(\brho)=
\left( 1+\frac{\Lambda^2}{K_0} \right) \brho -\sqrt{\frac{2d \Lambda^2}{K_0} 
 \brho(\brho-\brho_{\rm cl})}
$$
\begin{figure}[t]
\begin{center}
\vspace*{.1cm}
{\includegraphics[width=6.7cm]{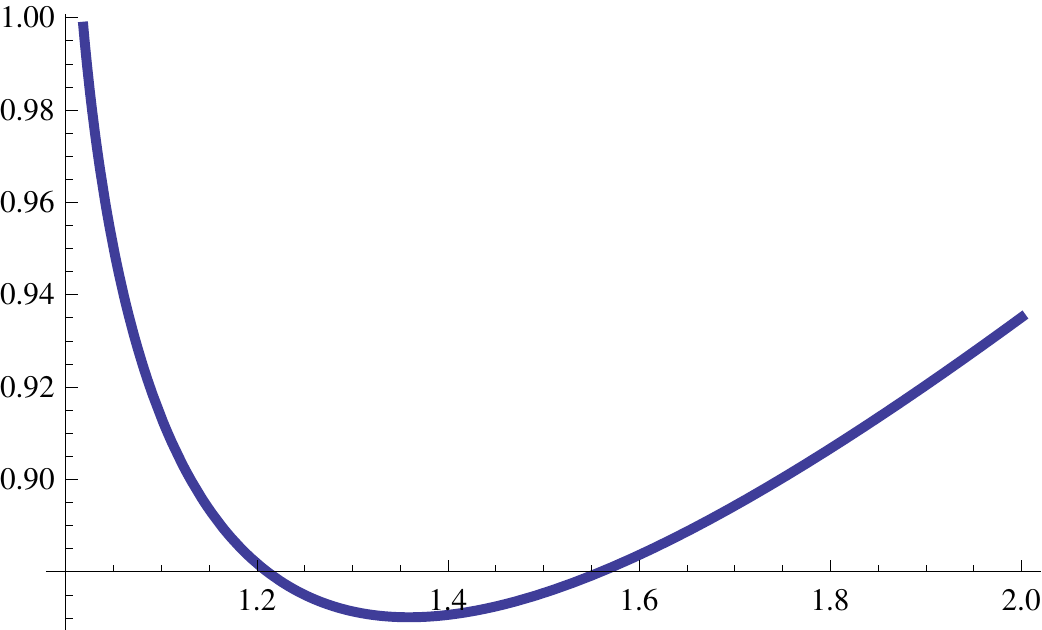}} 
\caption{Plot of the effective potential $\Gamma[\brho]$ illustrating an instability of
the classical ground state at $\brho=\brho_{\rm cl}$. 
The minimum is reached at the mean-field ground state.}
\end{center}
\labelf{fig01}
\vspace{-5mm}
\end{figure}
\!\!It is plotted in Fig.~\ref{fig01} versus  $\brho /\brho_{\rm cl}$.
The classical  ground state $\brho = \brho_{\rm cl}$ 
(at the left end) is \tcy{unstable} and a
\tcy{stable} minimum occurs at the mean-field value  (the same as before) if $K_0>K_* $. 

In the Lilliputian scaling regime  $\Gamma[\brho]$ is finite 
near the minimum and is given by 
a quadratic form which is positive defined,
illustrating the \tma{global stability} under fluctuations.
The nonlinearity results in  the \tbl{string susceptibility}
\fbox{$\gamma_{\rm str}=1/2$}  for a \tbr{cylinder} and a \tbr{torus}  
as shown by {\it Ambj\o rn, Y.M.\ (2017), (2021)} which
is quite different from \fbox{$\gamma_{\rm str}=1$} of KPZ-DDK.

\section{{\bf Fluctuations about  the mean field}\label{s:6}}

Expanding  the \tbl{effective action} about the mean-field ground state,
$\lambda^{ab}=\blambda \delta^{ab}+\delta \lambda^{ab}(\omega)$
and  $\rho=\brho+\delta \rho(\omega)$, we observe
the {same stability of quadratic wavy fluctuations  as about
the classical ground state
because of the background independence.} We have a
\tre{positive definite} quadratic form for \tma{imaginary} $\delta \lambda^{ab}$ and
\tma{real} $\delta \rho$. Again typical $\delta\lambda\sim1/\Lambda$
so $\lambda^{ab}$ is \tbl{localized}.
Thus \tre{only $\rho$ propagates} to macroscopic distances and its smooth fluctuations
are described by the usual Liouville action which is stable for $2<d<26$. 

This is not however the whole story because of
the private life of the fluctuating fields that
occurs at the distances $\sim \Lambda^{-1}$ but is  nevertheless observable as
demonstrated by
{\it {Y.M. (2021)}}. I shall now describe this issue.

\tit{Path integrating over $\lambda^{ab}$}

Let us set $\lambda^{z\bar z}=0$ and consider a simplified action ($\rho= \brho \e^{\vp}$)
$$
{\cal S}= \int \left[\frac 1{4\pi b^2_0} \partial \vp \bar \partial \vp + \left(
\lambda^{zz} \nabla \partial   \vp
+\lambda^{\bar z \bar z} \bar \nabla\bar \partial  \vp\right)- d{\Lambda^2 \brho}
\e^{\vp} \lambda^{zz} \lambda^{\bar z \bar z}\right]\!,
~~~ b_0^2=\frac 6{26-d}
$$
where the last term with $\Lambda^2$ illustrates the statement of the previous paragraph
about the localization.
Integrating out $\lambda^{zz} $ and $\lambda^{\bar z \bar z}$
and integrating by parts, we find  (only these two terms are independent)
$$
{\cal S} 
=\frac 1{4\pi b_0^2}\int \left\{\partial \vp \bar \partial \vp + 4\eps\e^{-\vp}\left[
(\partial\bar\partial \vp)^2 +\tbl{\partial \vp\bar\partial \vp \partial\bar\partial \vp}
\right]\right\}  , \quad \eps\propto\Lambda^{-2}
$$
modulo \tcy{boundary terms}.
The first additional term on the right-hand side 
 appears already
for \tgr{Polyakov's} string from the \tgr{Seeley} expansion of the heat kernel but the 
\tbl{second}  does \tre{not}.

\tit{Higher-derivative action}

Thus, integrating over $X^\mu$, ghosts, Pauli-Villars' 
regulators 
and 
$\lambda^{ab}$, we expect for the Nambu-Goto string
 the \tbl{beyond Liouville action}
$$
{\cal S}=\frac 1{16 \pi b_0^2} \int \sqrt{g} \left[g^{ab} \p_a \vp \p_b \vp  +2m_0^2
+\eps  R \left(R+\tbl G g^{ab}\,\partial_a \vp\partial_b \vp \right)\right]  
$$
Here $\vp=-\Delta^{-1} R$ becomes a local field in the conformal gauge.  
Once again, the $R^2$ appears already for the Polyakov  string but the second (nonlocal)  term 
with $\tbl G\neq 0$ is specific to the Nambu-Goto string.

Of course the higher-derivative terms are negligible classically for smooth metrics with 
$\eps R\ll1$, reproducing the Liouville action. However, the
quartic derivative provides both a UV cutoff and also an \tre{interaction} whose
\tbl{coupling} constant is $\eps$.  We thus encounter 
 uncertainties like $\eps \times \eps^{-1}$ so the higher-derivative terms  revive \tma{quantumly}. In other words the
\tbr{smallness of $\eps $ is compensated by a change of  the metric (the shift of $\vp$)}
what is specific to the theory with diffeomorphism invariance.

 The described procedure looks like an appearance of anomalies in quantum field
 theory. We may expect for this reason that possible yet higher-derivative terms 
 will not change the results. An argument in favor of such a universality 
 at $G=0$ was given by
 {\it Y.M.~(2021)}.

\section{\bf {Comparison with KPZ-DDK}}

\tit{Energy-momentum tensor}  

It is instructive to begin with the energy-momentum tensor of a scalar minimally coupled
to gravity
\bea
&&-4 b_0^2 T_{ab}^{({\rm min})}=\p_a \vp \p_b \vp -\frac 12 g_{ab} \p^c \vp \p_c \vp
 -\mu_0^2 g_{ab}-\eps \p_a \vp \p_b \Delta \vp -
\eps \p_a \Delta \vp  \p_b \vp  \non&& \hspace*{12mm}
+\eps g_{ab} \partial^c \vp\p_c \Delta \vp+\frac \eps2 g_{ab} (\Delta \vp)^2 
-G \eps \p_a \vp\p_b\vp \Delta \vp 
 + G\frac\eps2\p_a\vp \p_b(\p^c \vp \p_c \vp) \non &&\hspace*{12mm}
+ G\frac\eps2 \p_a (\p^c \vp \p_c \vp)\p_b\vp -G\frac\eps2
g_{ab} \p^c\vp \p_c (\p^d \vp \p_d \vp)
\nonumber
 \eea
The result by {\it Gibbons, Pope, Solodukhin (2019)} is reproduced {at $G=0$}.

Additional terms emerge 
 for our \tma{diffeomorphism invariant} action so the energy-momentum tensor 
as derived by {\it Y.M.~(2022)} reads 
 \bea
-4 b_0^2 T_{ab} &=&-4 b_0^2 T_{ab}^{({\rm min})}
-2 (\p_a\p_b-g_{ab}\p^c\p_c )(\vp -\eps \Delta \vp+G\frac\eps2 g^{ab}\p_a \vp \p_b \vp  ) 
\non &&
 + \tbl{2 G \eps (\p_a \p_b-g_{ab} \p^c\p_c) 
 \frac 1{\Delta} \partial^d(\p_d \vp \Delta\vp) }\nonumber
\eea
It is \tre{conserved} and \tre{traceless} (\tre !) thanks to diffeomorphism invariance
in spite of $\eps$ is dimensionful. Notice the nonlocality of the last term inherited from
nonlocality of the covariant action.

The
$T_{zz}$ component
  \bea
  \lefteqn{
 -4 b_0^2 T_{zz}=(\p \vp)^2 -2\eps \p \vp \p \Delta \vp -2\p^2 (\vp - \eps\Delta \vp)
-G \eps (\p \vp)^2 \Delta \vp} \non
 && + G\eps \Big[4\p\vp \p(\e^{-\vp}\p \vp \bp \vp) 
- 4 \p^2(\e^{-\vp}\p \vp \bp \vp)
 +  \partial(\p \vp \Delta\vp)
  +  \tbl{ \frac 1{\bp}\p ^2 (\bp \vp \Delta\vp) } \Big]\nonumber
  \eea
 reproduces in two dimensions {at $G=0$} the one by 
{\it Kawai, Nakayama (1993)}. 

\tit{DDK for the beyond Liouville action} 
 
 Given $T_{zz}$ it is possible to perform \'a la DDK the computation of the conformal weight and the central charge at one loop about either the clasical of mean-field ground states.
The results are the same because of the background independence.
The operator products $T_{zz}(z) \e^{\vp(0)}$ and  $T_{zz}(z) T_{zz}(0)$ 
are given at one loop by a bunch of diagrams most of which can be described by introducing 
\begin{figure}[t]
\begin{center}
\vspace*{.2cm}
\includegraphics[width=6.7cm]{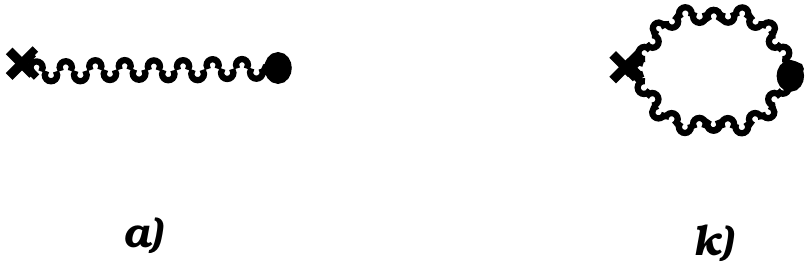} 
\caption{Diagrams for  the operator products 
$T_{zz}(z) \e^{\vp(0)}$ or $T_{zz}(z) T_{zz}(0)$. 
}
\end{center}
\labelf{fi:2}
\vspace{-5mm}
\end{figure}
an effective $T_{zz}$ for smooth fields similarly to DDK 
$$
-4 b^2T_{zz}^{({\rm eff})}=
(\partial\varphi )^2 -2q \partial^2 \varphi 
$$


For the conformal weight of $\e^{\vp(0)}$, only this effective $T_{zz}$  contributes as $\eps\to0$
so we have $1=\tbr{q-b^2}$  as usual.
For the central charge we have usual \tbr{$6 q^2/b^2+1$}  from 
$ T_{zz}^{({\rm eff})}$ but now
the \tbl{nonlocal} term in $T_{zz}$ revives in the diagram k) and gives additionally $6 G $.
I shall momentarily return to this most interesting result.
There is also a logarithmically divergent term whose appearance I link to
the subtleties with conformal Ward identities for $\tbl{G}\neq0$
because $T_{zz}$ is then not primary.


\section{{\bf Algebraic check of DDK}}

\tit{Pauli-Villars' regularization}  

To describe the one-loop renormalization in the standard way, we add
{Pauli-Villars' regulators}: \tbr{Grassmann} $Y$, $\bar Y$ (of mass squared $M^2$) and 
\tbr{normal} $Z$ (of mass squared $2M^2$) as proposed by {\it Ambj\o rn, Y.M.\ (2017)} at $\eps=0$.
This regularizes all involved divergences.
To simplify formulas I keep below only $Y$ which is enough to compute finite parts.
The regulator action  reads
$$
{\cal S}_{\rm reg.} =\frac 1{16\pi b_0^2}\int \sqrt{g}\left[ g^{ab}\partial_a  Y \partial_b Y +M^2 Y^2+
\eps (\Delta Y)^2 +G\eps g^{ab}\partial_a  Y \partial_b Y R  \right] 
$$

The regulators makes a contribution to the energy-momentum tensor
which is quadratic in the regulator fields and local. 
 The total energy-momen\-tum tensor is
 \tcy{conserved} and \tcy{traceless} (\tre !) in spite of the masses.
 We thus expect conformal invariance  to be \tma{maintained quantumly} what can be
 explicitly demonstrated by the one-loop and some two-loop calculations.
 

\tit{Explicit one-loop results}

The
 renormalization of  $b^2$  comes from the usual one-loop diagrams including tadpoles
$$
\frac{1}{b^2}=\frac1{b_0^2}-
\left(\frac 16 -4 +A +2 G \int \d k^2 \frac \eps{(1+\eps k^2)}-
\frac 12 G A\right)+{\cal O}(b_0^2) 
$$
Here $A(\eps M^2)\sim \sqrt{\eps} M$ is the contribution of the tadpole. 

The analogous 
one-loop renormalization of $q/b^2$ reads
$$
\frac q{b^2}=\frac{1}{b_0^2}-\frac 16 +2 -\frac 12 A -\frac 12 G 
- G \int \d k^2 \frac {\eps }{(1+\eps k^2)} +\frac 14 G A +{\cal O}(b_0^2)
$$
or, multiplying by $6 b^2$,
$$
\frac {6q^2}{b^2} =\left(\frac q{b^2}\right)^2\times 6 b^2
=\frac{6}{b_0^2} - 1 -6 G +{\cal O}(b_0^2)
$$
This precisely \tre{confirms} the above shift of the central charge by $\tbl{6G}$
obtained by the \tcy{conformal field theory} technique of \tgr{DDK}.

\section{{\bf Conclusion}}


The classical (perturbative) ground state of the Nambu-Goto string is stable only for $d<2$.
For $2<d<26$ the \tre{mean-field} ground state is stable instead and
we have the 
{Lilliputian strings} for $d>2$ versus Gulliver's strings  for $d\leq 2$.
{Higher-derivative terms} in the \tbr{beyond Liouville action} for $\vp$ revive,
telling the \tgr{Nambu-Goto} and \tgr{Polyakov} strings apart.
Two-dimensional \tma{conformal invariance} is \tre{maintained} by fluctuations in spite of 
dimensionful \tbr{$\eps$}
 but the \tre{central charge} of $\vp$ gets \tbl{additional $6 G$} at one loop.

All this is specific to the theory with \tma{diffeomorphism invariance}.
My final remark is about yet another diffeomorphism invariant theory -- Gravity.
The
\tcy{large-$d$ strings are described by bubble diagrams like the $O(N)$ sigma model but the
large-$d$ gravity is described by planar diagrams like Yang-Mills 
as pointed out by {\it {Strominger (1981)}}}.
This is the next level of complexity.

This work was supported by the Russian Science Foundation (Grant No.20-12-00195).


\end{document}